\documentclass[aps,prx,floatfix,twocolumn,superscriptaddress]{revtex4-2}
\usepackage{graphicx} \usepackage{subfigure}
\usepackage{stdclsdv} \usepackage{array}  \usepackage{amsmath}
\usepackage{amsmath}
\usepackage{siunitx}
\usepackage{amsfonts}
\usepackage{amssymb}
\usepackage{xcolor}
\usepackage{float}

\usepackage{hyperref}
\usepackage{xcolor}
\hypersetup{
    colorlinks,
    linkcolor={red!50!black},
    citecolor={blue!50!black},
    urlcolor={blue!80!black}
}

\newcommand{\Fig}[1]{Fig.~\ref{#1}}
 
\newcommand{\Sec}[1]{Sec.~\ref{#1}}

\newcommand{\App}[1]{SI Sec.~\ref{#1}}
\newcommand{\Apps}[1]{SI Secs.~\ref{#1}}
\newcommand{\Eq}[1]{Eq.~(\ref{#1})}
\newcommand{\Eqs}[1]{Eqs.~(\ref{#1})}

\begin{document}

\title{ Liquid demixing in elastic networks: cavitation, permeation,  or size selection? }
 
\author{Pierre Ronceray}
\email{pierre.ronceray@univ-amu.fr}
\affiliation{Center for the Physics of Biological Function, Princeton University, Princeton, New Jersey 08544, USA}
\affiliation{Aix Marseille Univ, Universit\'e de Toulon, CNRS, CPT, Turing Center for Living Systems, Marseille, France}

\author{Sheng Mao}
\affiliation{Department of Mechanics and Engineering Science, BIC-ESAT, College of Engineering, Peking University, Beijing 100871, People's Republic of China}
\affiliation{Department of Mechanical and Aerospace Engineering, Princeton University, Princeton, New Jersey 08544, USA}

\author{Andrej Ko\v{s}mrlj}
\affiliation{Department of Mechanical and Aerospace Engineering, Princeton University, Princeton, New Jersey 08544, USA}
\affiliation{Princeton Institute for the Science and Technology of Materials (PRISM),
Princeton University, Princeton, New Jersey 08544, USA}

\author{Mikko P. Haataja}
\email{mhaataja@princeton.edu}
\affiliation{Department of Mechanical and Aerospace Engineering, Princeton University, Princeton, New Jersey 08544, USA}
\affiliation{Princeton Institute for the Science and Technology of Materials (PRISM),
Princeton University, Princeton, New Jersey 08544, USA}

\maketitle

\textbf{Demixing of multicomponent biomolecular systems via
  liquid-liquid phase separation (LLPS) has emerged as a potentially
  unifying mechanism governing the formation of several membrane-less
  intracellular organelles
  (``condensates'')~\cite{brangwynne_germline_2009,alberti_considerations_2019,hyman_liquid-liquid_2014,berry_physical_2018,bracha_probing_2019,choi_physical_2020,feric_coexisting_2016},
  both in the cytoplasm (\emph{e.g.}, stress granules) and in the
  nucleoplasm (\emph{e.g.}, nucleoli). While both \emph{in vivo}
  experiments \cite{shin_liquid_2018} and studies of synthetic systems
  \cite{style_liquid-liquid_2018,rosowski_elastic_2020} demonstrate
  that LLPS is strongly affected by the presence of a macromolecular
  elastic network, a fundamental understanding of the role of such
  networks on LLPS is still lacking.  Here we show that, upon
  accounting for capillary forces responsible for network expulsion,
  small-scale heterogeneity of the network, and its nonlinear
  mechanical properties, an intriguing picture of LLPS
  emerges. Specifically, we predict that, in addition to the
  experimentally observed cavitated droplets
  \cite{shin_liquid_2018,style_liquid-liquid_2018} which fully exclude
  the network, two other phases are thermodynamically possible:
  elastically arrested, size-limited droplets at the network pore
  scale, and network-including macroscopic droplets. In particular,
  pore size-limited droplets may emerge in chromatin networks, with
  implications for structure and function of nucleoplasmic
  condensates.}

\begin{figure}[b]
  \centering
  \includegraphics[width=\columnwidth]{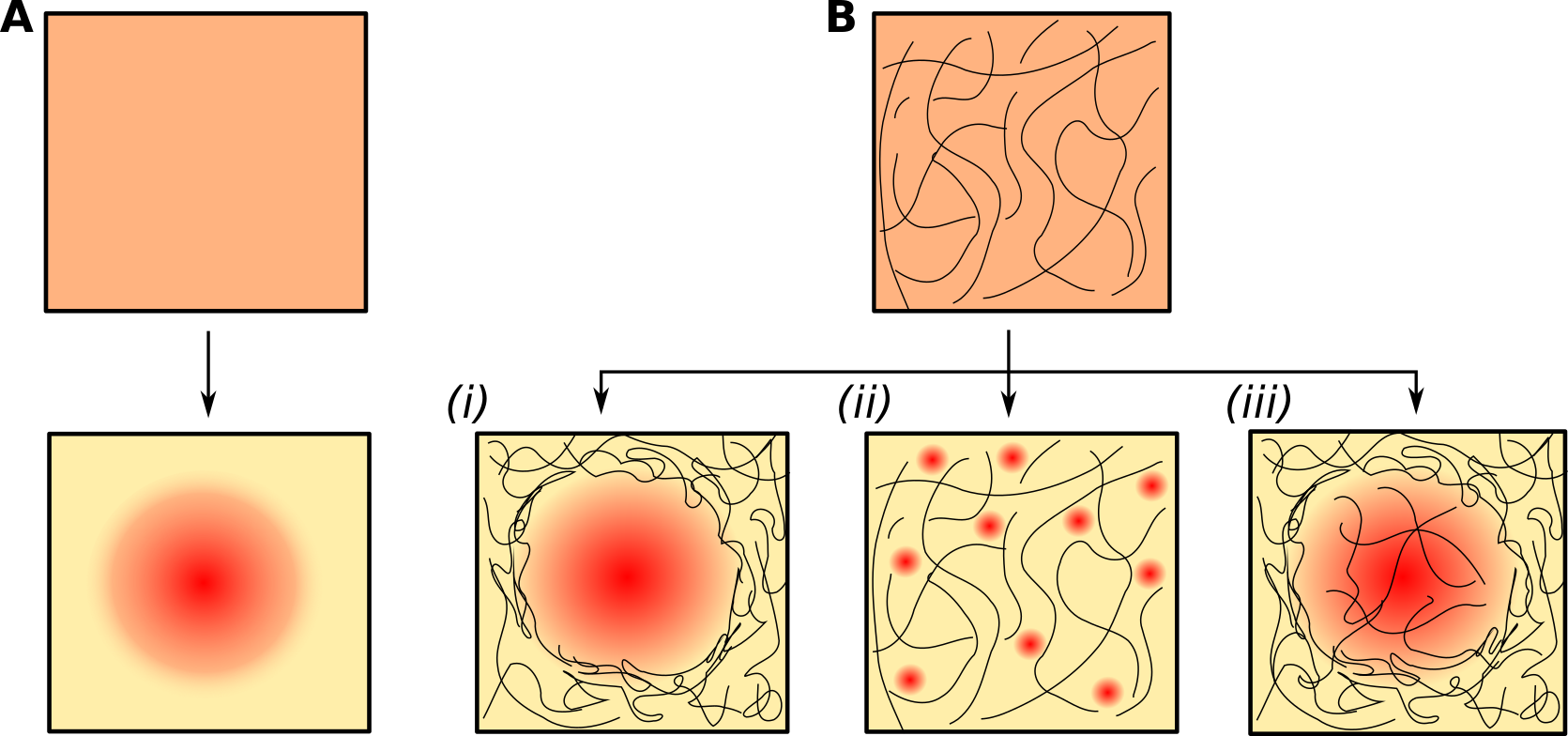}
  \caption{ \textbf{A.}~Liquid-liquid phase separation (LLPS) from an
    initially mixed phase (\emph{top}) results in a macroscopic
    droplet of the red minority liquid immersed in the yellow majority one
    (\emph{bottom}). \textbf{B.}~When LLPS occurs with the initial
    mixed phase imbibed in an elastic network (\emph{top}), we
    identify three possible outcomes (\emph{bottom}): \emph{(i)}
    Cavitation. \emph{(ii)} Microscale droplets. \emph{(iii)}
    Permeation of the network into the minority phase.  }
  \label{fig:scenarios}
\end{figure}

When LLPS occurs without mechanical constraints
(\Fig{fig:scenarios}A), the thermodynamically stable outcome of the
demixing is a macroscopic spherical droplet of the minority liquid
(red) embedded within the majority phase (yellow). This ensures that
the contact surface per droplet volume between the two liquids is
minimized, and results in a negligible, sub-extensive free energy
penalty compared to the bulk phase-separated liquid. In the presence
of an elastic matrix hindering LLPS, in contrast, we distinguish and
study here three distinct scenarios by which demixing can occur
(\Fig{fig:scenarios}B). Each scenario results in a specific free
energy cost compared to the reference, unhindered case: \emph{(i)} The
minority liquid can create a macroscopic cavity, which incurs a
deformation energy penalty $E_\mathrm{el}$ associated with the elastic
matrix. \emph{(ii)} Alternatively, the minority liquid may form an
extensive number of microdroplets fitting within the pores of the
network, which avoids elastic deformation but incurs an extensive
surface energy penalty $E_\mathrm{surf}$. \emph{(iii)} Finally, rather
than fully excluding the network, the minority droplet can permeate
through it, resulting in a finite wetting energy $E_\mathrm{wet}$
between the droplet and the network. Below, we establish an
equilibrium phase diagram for LLPS within an elastic network by
assessing the relative thermodynamic stability of each scenario.

Specifically, we introduce physically-based models for each of the
three energy penalties compared to the reference situation of demixing
in the absence of a network (\Fig{fig:scenarios}A), and evaluate the
stability of each phase by quantifying this free energy cost per
volume of phase-separated minority liquid, an approach previously used
fruitfully in the context of block copolymer phase
behavior~\cite{semenov_contribution_1984}. We focus on identifying the
thermodynamically stable droplet configurations, and
thus ignore all kinetic processes (incl.~nucleation, growth and
coarsening). In addition, we assume that the droplets occupy a
negligible volume fraction, and hence neglect all elastic and chemical
inter-droplet interactions.  We first employ simple scaling arguments
to establish a morphological phase diagram by considering only the
dominant term(s) of the free energy for each phase. Motivated by
existing theoretical
approaches~\cite{style_liquid-liquid_2018,rosowski_elastic_2020,kim_extreme_2020,kothari_effect_2020},
we then develop a comprehensive theory of droplets constrained by
elastic networks by accounting for capillary forces responsible for
network expulsion from droplets, as well as heterogeneities in the
network structure and its nonlinear mechanical properties. Both
analytical and numerical approaches are employed to confirm the
salient features of the phase diagram and elucidate the nature of phase transitions between the droplet phases.

We begin by considering scenario \emph{(i)}, where a macroscopic
droplet of size $r\to\infty$ forms by creating a network-excluding
cavity. This scenario was previously considered for \emph{in vitro}
oil-water mixtures in silicone
gels~\cite{style_liquid-liquid_2018,rosowski_elastic_2020,kim_extreme_2020}
and \emph{in vivo} droplets in the cell
nucleus~\cite{shin_liquid_2018}. In order to form a macroscopic cavity
from an initial pore, large deformations must occur in the
network. Therefore, it is necessary to go beyond simple linear
elasticity in the treatment of the network mechanics. The simplest
such extension is a neo-Hookean (NH) constitutive relation, considered
in
Refs.~\cite{shin_liquid_2018,style_liquid-liquid_2018,rosowski_elastic_2020},
where the elastic energy
\begin{equation}
  \label{eq:NH_scaling}
  E_\mathrm{el}(r) \sim \frac{4\pi r^3}{3} \alpha G
\end{equation}
scales as the volume of the cavity when $r\to\infty$. Here, $G$
denotes the shear modulus of the network, while the numerical
coefficient $\alpha \sim \frac{5}{2}$ is a material parameter.  This simple behavior reasonably describes a broad class of
artificial gels~\cite{treloar_physics_2005}, and several mechanisms
can lead to such a volume scaling, such as detachment of cross-links
or fracture at fixed hoop stress
\cite{raayai-ardakani_intimate_2019,vidal-henriquez_cavitation_2021}. For
this scenario, the free energy per volume penalty compared to the
reference system without a network is thus
\begin{equation}
  \label{eq:Delta_cavitation}
  \Delta g_{(i)} \sim \alpha G.
\end{equation}
This constant free energy penalty results in a shift of the phase
boundary to lower temperatures. Remarkably, this behavior was
characterized and validated for \emph{in vitro}
systems~\cite{style_liquid-liquid_2018}, with a value
$\alpha \approx 1.5$. In the presence of macroscopic gradients in the
network stiffness, \Eq{eq:Delta_cavitation} also implies that droplet
growth is favored in softer regions of the system, where the
phase-separated liquid has a lower free
energy~\cite{shin_liquid_2018,rosowski_elastic_2020,vidal-henriquez_theory_2020,rosowski_elastic_2020-1}.

\begin{figure}[b]
  \centering
  \includegraphics[width=0.8\columnwidth]{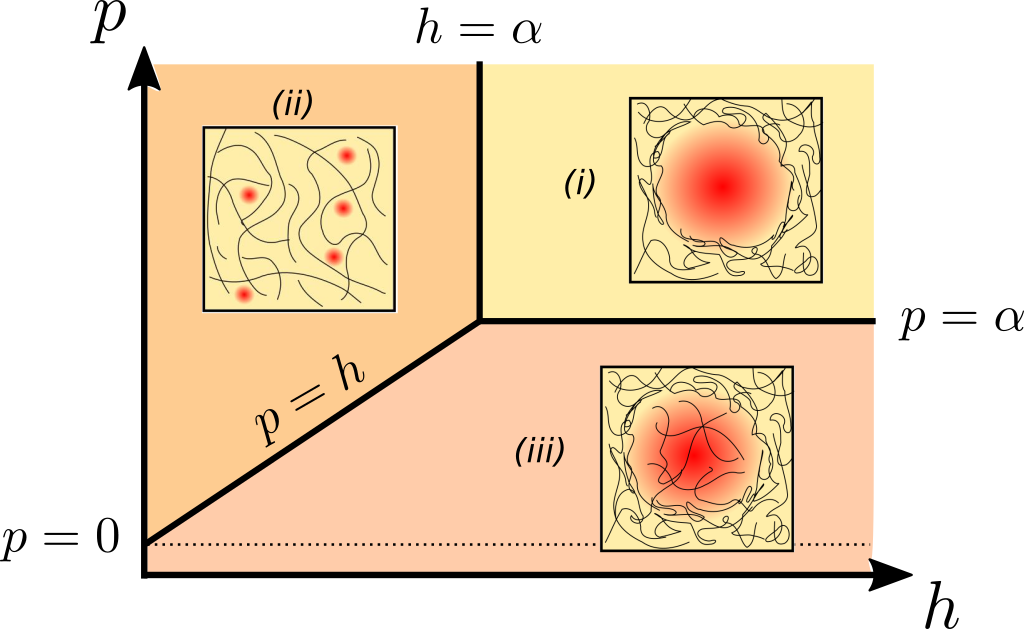}
  \caption{ Putative phase diagram from simple scaling arguments, indicating the most stable
    state for liquid-liquid phase separation in an elastic network, as
    a function of the elasto-capillary number $h$ (\Eq{eq:h}) and the
    permeo-elastic number $p$ (\Eq{eq:p}). Note that only the dominant
    contribution to the free energy is retained here, corresponding to
    Eqs. (\ref{eq:Delta_cavitation}), (\ref{eq:Delta_microdroplets}) and
    (\ref{eq:Delta_permeated}), respectively for the cavitated
    (\emph{i}), micro-droplets (\emph{ii}) and permeated (\emph{iii})
    phases.}
  \label{fig:PD}
\end{figure}

While this model captures the macroscopic elastic response of the
material, it does not account for small-scale heterogeneities. In both
biological and artificial systems considered here, the elastic network
is constituted by polymers with a finite pore size $\xi$
characterizing the size of interstices between polymers. Consider now scenario
\emph{(ii)} in \Fig{fig:scenarios}B, in which microdroplets with
$r=\xi$ form within these pores without deforming the network. In this case, $E_\mathrm{el} = 0$, while due to their small size, the droplets incur a substantial surface energy penalty
$E_\mathrm{surf} = 4\pi \xi^2 \gamma$, where $\gamma$ denotes the surface
tension between the two liquid phases. Per volume of the minority species, this
result in a free energy penalty
\begin{equation}
  \label{eq:Delta_microdroplets}
  \Delta g_{(ii)} \sim \frac{3 \gamma}{\xi}
\end{equation}
for scenario \emph{(ii)}, compared to our reference system in absence of elastic
network. Comparing \Eqs{eq:Delta_microdroplets} and
(\ref{eq:Delta_cavitation}) reveals that in such a porous network, the
trade-off between elastic and surface energy is controlled by the \emph{elasto-capillary
  number}~\cite{shao_elastocapillary_2019}:
\begin{equation}
  \label{eq:h}
  h \equiv \frac{ 3\gamma }{\xi G}.
\end{equation}
When $h > \alpha$, \emph{i.e.}, for an elastically homogeneous network
and large liquid-liquid surface tension, scenario \emph{(i)} is
thermodynamically favored, leading to the formation of macroscopic
cavitated droplets. In contrast, when $h < \alpha$, pore-size-limited
microdroplets corresponding to scenario \emph{(ii)} are thermodynamically more
stable (\Fig{fig:PD}). 

\begin{figure*}[bt]
  \centering
  \includegraphics[width=1.0\textwidth]{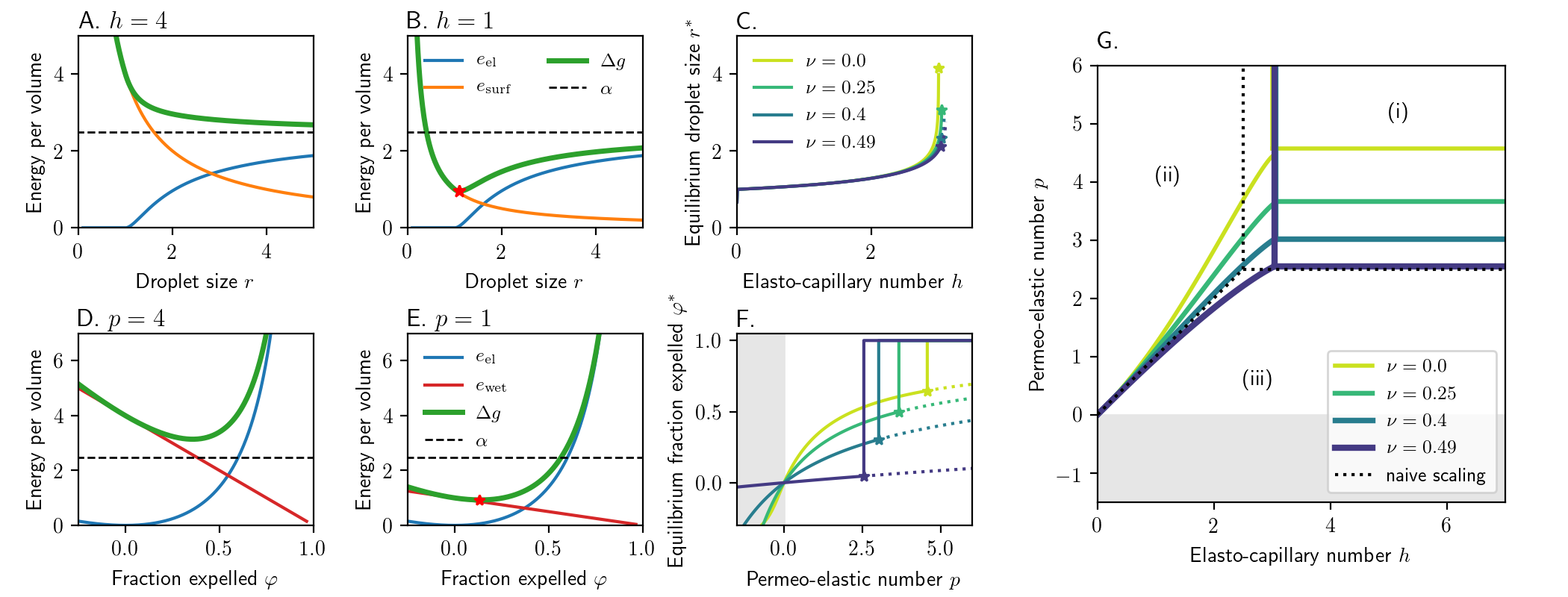}
  \caption{ Analysis of droplet phases within compressible NH
    networks. \textbf{A-B.}  Elastic energy (blue, $e_\text{el}$),
    surface energy (orange, $e_\text{surf}$) and total free energy
    (green, $\Delta g=e_\text{el}+e_\text{surf}$) per volume for a
    droplet as a function of pore size $r$, respectively for
    elasto-capillary numbers $h=4$ (showing monotonic decay of
    $\Delta g$) and $h=1$ (showing a global minimum of
    $\Delta g$ at $r^*$, red
    star). Dashed black line indicates the $\lambda\to\infty$
    cavitated limit. \textbf{C.} Equilibrium pore size $r^*$ as a
    function of the elasto-capillary number $h$, for different Poisson
    ratios $\nu$ of the network. Star indicates the limit of stability
    of phase \emph{(ii)}. \textbf{D-E.} Elastic, wetting (red,
    $e_\text{wet}$) and free energy (green,
    $\Delta g = e_\text{el}+e_\text{wet}$) per volume of a large
    droplet permeating through the network, as a function of the
    expelled volume fraction $\varphi$ of the network, respectively
    for permeo-elastic numbers $p=4$ (where cavitation is favored) and
    $p=1$ (with global minimum at $\varphi^*$, red star). \textbf{F.}
    Equilibrium expelled volume fraction $\varphi^*$ of the network as
    a function of the permeo-elastic number $p$. Dashed lines indicate
    metastable states, with cavitation ($\varphi^*=1$) energetically
    favored. \textbf{G.}  Phase diagram indicating the most stable
    phase in the $(p,h)$ plane. Dotted lines indicate naive scaling
    results with $\alpha=5/2$, as in \Fig{fig:PD}. The shaded area in
    F,G indicate $p<0$, \emph{i.e.}, a contractile droplet attracting
    the network. In A,B,D,E we take $\nu = 0.4$. Energy densities and
    length scales are respectively normalized by the linear shear
    modulus $G$ and the pore size $\xi$. }
  \label{fig:NH}
\end{figure*}

In the scenarios considered thus far, the network is fully excluded
from the droplets. We now consider scenario \emph{(iii)} from
\Fig{fig:scenarios}B: the partial inclusion of the network in
macroscopic droplets of the minority phase. To assess the stability of
this scenario, we introduce a wetting energy
$E_\mathrm{wet} $, emanating from the minority phase permeating
through the network, as
\begin{equation}
  \label{eq:Ewet}
  E_\mathrm{wet} = \frac{4\pi r^3}{3} (1-\varphi) \sigma_p,
\end{equation}
where $\varphi$ denotes the fraction of network expelled from the
droplet compared to the undeformed state, and $\sigma_p$ denotes the
\emph{permeation stress}.  Microscopically, $\sigma_p$ arises from
differential wetting energy per unit length of filaments constituting
the network in contact with the two
fluids~\cite{de_gennes_liquid-liquid_1984} (see \App{sec:permeation}).
\Eq{eq:Ewet} translates this microscopic wetting phenomenon into a
macroscopic effect, which results in a stress discontinuity at the
liquid-liquid interface through which the network permeates. In
addition to a bulk energy term (\Eq{eq:Ewet}), network wetting can
induce an effective change of liquid-liquid surface energy, in
particular if filaments align with the interface. We do not consider
such an effect in this article.

Again, in the spirit of a simple scaling analysis,
we first neglect the network deformation in response to this stress
and set $\varphi = 0$. The free energy per volume corresponding to
this permeated scenario is thus
\begin{equation}
  \label{eq:Delta_permeated}
  \Delta g_{(iii)} \sim \sigma_p.
\end{equation}
Comparing this expression with \Eq{eq:Delta_cavitation}, we find that
the most stable phase is controlled by a second dimensionless
quantity, namely the \emph{permeo-elastic number}
\begin{equation}
  \label{eq:p}
  p \equiv \frac{\sigma_p}{G},
\end{equation}
which is a measure of the degree of network deformation at the
interface induced by the permeation stress. For $p>\alpha$, scenario
\emph{(i)} is the most stable: the repulsion between the network and
the minority liquid is sufficiently strong to fully expel the network
from the droplet, leading to cavitation. For $p<\alpha$, the droplet
permeates through the network rather than excluding it, and scenario
\emph{(iii)} is preferred.  Finally, when the elasto-capillary number
$h<\alpha$, the phase boundary between scenarios \emph{(ii)} and
\emph{(iii)} is given by the line $p=h$.

The results of this scaling analysis are summarized in a phase diagram
in the $(p,h)$ plane in \Fig{fig:PD}, which predicts the most stable
demixed phase. These phase boundaries depend only on the liquid and network properties, not on the degree of supersaturation:
to assess whether demixing takes place or not, the free energy penalty
of the most stable phase (Eq.~\ref{eq:Delta_cavitation},
\ref{eq:Delta_microdroplets} or \ref{eq:Delta_permeated}) should be
added to the demixing free energy per volume in the absence of
network. We note that for scenarios \emph{(i-ii)}, the network hinders
phase separation and stabilizes the mixed phase; for scenario
\emph{(iii)}, this depends on the sign of $p$: for $\sigma_p<0$, the
network prefers the minority phase and favors phase separation.

We have so far considered only the dominant contribution to the free
energy for each scenario -- either $E_\mathrm{el}$, $E_\mathrm{surf}$
or $E_\mathrm{wet}$. Network deformation will however occur in each of
the three scenarios: in \emph{(ii)}, microdroplets exert a pressure on
the network, while in \emph{(iii)}, a permeation stress $\sigma_p>0$
results in a partial expulsion of the network from the droplet. To
quantitatively predict the locations of the phase boundaries between
scenarios and the nature of associated phase transitions, we next
discuss the deformation behavior arising from an isolated droplet
embedded within a slightly compressible NH network
(see~\Apps{sec:framework}-\ref{sec:NH}).

Examining first scenarios \emph{(i-ii)} for which the network is fully
excluded from the droplet, we consider a droplet of radius $r$ in a
spherical cavity of initial radius $\xi$ that corresponds to the
characteristic pore size of the network. When the elasto-capillary
number $h$ is large (\Fig{fig:NH}A), the free energy per volume of the
droplet $\Delta g = e_\mathrm{el}+e_\mathrm{surf}$ decreases
monotonically with droplet size $r$, indicating that cavitation
(scenario \emph{i}) is thermodynamically favored.  At small $h$
(\Fig{fig:NH}B), in contrast, the free energy exhibits a global
minimum at $r^*\gtrsim \xi$, and size-limited microdroplets with
radius $r^*$ as per scenario \emph{(ii)} are favored. For positive
Poisson's ratios $\nu$, the radius $r^*$ increases sharply with the
elasto-capillary number $h$ (\Fig{fig:NH}C), but remains finite up to
the limit of stability of microdroplets, indicating that the
cavitation transition \emph{(i$\to$ ii)} is weakly first-order as
surface tension is increased or, equivalently, as the shear modulus of
the network is reduced.  Interestingly, this transition becomes continuous
for auxetic materials with $\nu < 0$ (see~\App{sec:metastable}).

Turning now to the case of a permeated network with homogeneous
stretch $\lambda$ inside the droplet, we consider a macroscopic phase
separated droplet (thus neglecting the surface energy
$e_\mathrm{surf}$) for which the free energy per volume is a function
of the fraction $\varphi=1-\lambda^{-3}$ of the network expelled from
the droplet. When the permeo-elastic number $p$ is large, the free
energy exceeds that of the cavitated case for all $\varphi$
(\Fig{fig:NH}D). In contrast, at small values of $p$ (\Fig{fig:NH}E)
the global free energy density minimum occurs at a finite value
$\varphi^*$, and permeation is favored. When $p$ increases, the
equilibrium expelled network fraction $\varphi^*$ increases
continuously up to the cavitation point, at which it experiences a
compressibility-dependent jump (\Fig{fig:NH}F), implying that the
transition is discontinuous. We summarize these results in a phase
diagram for NH materials (\Fig{fig:NH}G).

\begin{figure}[bt]
  \centering
  \includegraphics[width=\columnwidth]{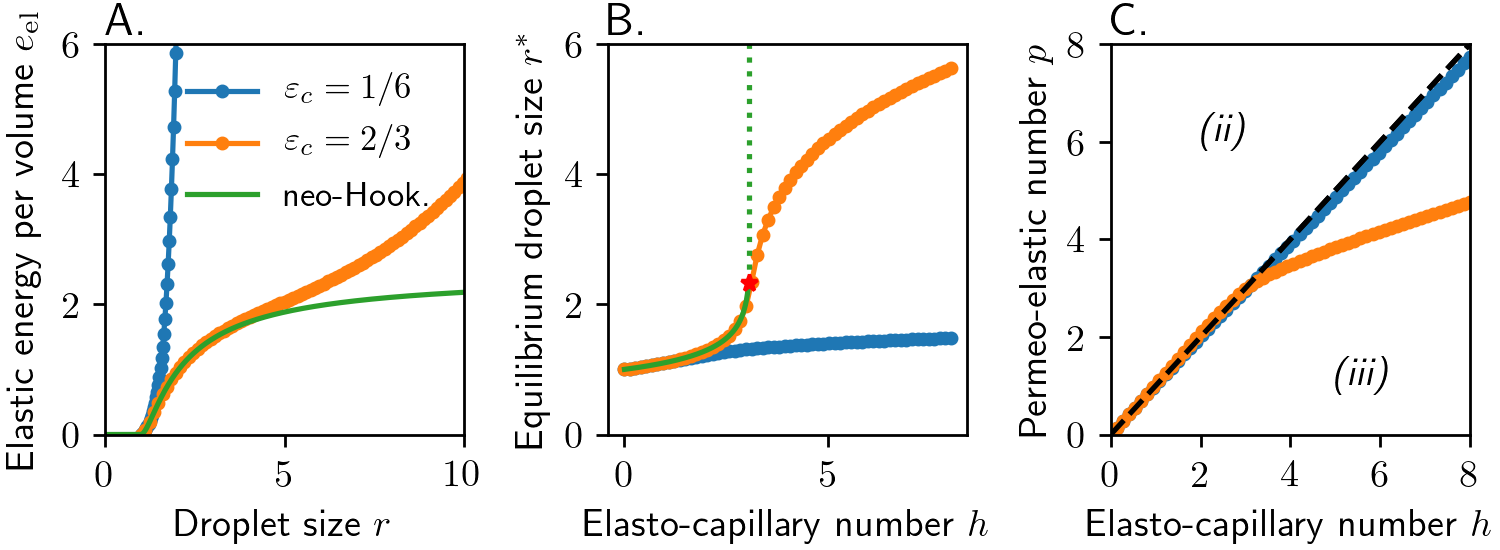}
  \caption{Numerical analysis of strain-stiffening materials with the
    nonlinear contribution to the elastic energy density described
    by a term $\propto ( (\lambda^2-1)/\varepsilon_c)^3$ (see \App{sec:NL}), where $\lambda$ is the stretch and
    the parameter $\varepsilon_c$ controls the strength of
    nonlinearity. Low (blue) and high (orange) values of
    $\varepsilon_c$ describe strong and weak nonlinearity,
    respectively. The analytical solution for non-stiffening
    NH materials is also shown (green).  \textbf{A.}
    Elastic energy per droplet volume as a function of droplet size.
    \textbf{B.}  Equilibrium droplet size $r^*$ as a function of the
    elasto-capillary number $h$. For NH materials the cavitation transition is shown as a dotted line. \textbf{C.}  Phase boundary between
    microdroplets \emph{(ii)} and permeated \emph{(iii)}
    phases. Cavitation \emph{(i)} is suppressed by the
    strain-stiffening. Dashed line indicates naive scaling $p=h$. }
  \label{fig:stiffening}
\end{figure}

While an NH constitutive law describes the deformation behavior of a
broad class of materials at finite stretches, many biomolecular networks differ by exhibiting nonlinear
\emph{strain-stiffening}
behavior~\cite{ogden_non-linear_1997,storm_nonlinear_2005,erk_strain_2010}
whereby the (nominal) tensile stress grows faster than linearly with
the stretch -- either as a power-law with exponent $>1$, or with a
divergence at finite stretch. In the permeated case, this nonlinearity
limits the exclusion of the network from the droplet, with moderate
effects on the phase stability. In contrast, strain stiffening
strongly affects phases \emph{(i-ii)} where the network is fully
excluded: the free energy of the cavity grows asymptotically faster
than its volume, and the elastic penalty $e_{\mathrm{el}}(r)$ diverges
in the limit of large droplets, as illustrated in
\Fig{fig:stiffening}A by numerical analysis of a minimal model for
power-law strain stiffening materials (see \App{sec:NL}). As a
consequence, effectively $\alpha\to\infty$, and scenario \emph{(i)} is
suppressed: the global energy minimum always occurs at a finite
droplet radius $r^*$, leading to size selection, as recently noted in
the context of the Gent model~\cite{wei_modeling_2020}. When the
nonlinearity is strong, the equilibrium droplet size is
$r^*\gtrsim \xi$ even at large capillary forces corresponding to
$h\gg 1$ (\Fig{fig:stiffening}B, blue), and microdroplets are stable
when $p\gtrsim h$ (\Fig{fig:stiffening}C). When the nonlinearity is
weak and emerges only at large stretch, in contrast, microdroplets
transition from being linearly arrested with size $r^*\gtrsim \xi$ at
$h \lesssim 3$, to being non-linearly arrested at a mesoscopic,
material dependent size $r^* \gg \xi$ at $h \gtrsim 3$
(\Fig{fig:stiffening}, orange). This transition is a smooth crossover
for realistic material parameters, and results in a change of slope in
the phase boundary between microdroplets and permeated droplets as
larger droplets incur a lower surface penalty (\Fig{fig:stiffening}C).

\begin{table*}[ptb]
  \caption{\label{tab:exps} Order-of-magnitude estimates of the shear
    modulus $G$, network mesh size $\xi$, surface tension $\xi$ and
    permeation stress $\sigma_p$ for three classes of experimental
    systems. We indicate the range of variation of the
    elasto-capillary number $h$ and the permeo-elastic number $p$, and
    conclude on the plausible scenarios for LLPS (most likely in
    bold). Details in \App{sec:exps}. }
\begin{tabular}{rl | c | c | c | c | c | c | c}
& System & $G$ & $\xi$ & $\gamma$ &$ \sigma_p$ & $h$ & $p$ & Scenarios \\
\hline
  I & Oil in silicone gel & $10^3 - 3.10^5 \si{\pascal}$ & $2-14 \si{\nano \meter}$ & \num{4e-3} \si{\newton\per\meter} & $10^4 - 3.10^5$ \si{\pascal} & 20-700 & 1.1 - 6.5 & \emph{\textbf{(i)}, (iii)}\\

    II & Cytoplasmic cond.  & $10 - 100$ \si{\pascal} & 50-150 \si{\nano \meter} & $10^{-6}$ \si{\newton\per\meter} & $\pm 0.2 - 2$ \si{\pascal} & \num{0.2}-\num{6} & $\pm 10^{-3} - 0.2$ & \emph{(ii), \textbf{(iii)}}\\

  III & Nuclear condensates & $10 - 10^3$ \si{\pascal} & 7-20 \si{\nano \meter} & $10^{-7}-10^{-6}$ \si{\newton\per\meter} & $\pm 10 - 100$ \si{\pascal} & \num{0.01}-\num{10} & $\pm 0.01-10$ & \emph{(i), (ii), (iii)}\\
\end{tabular}
\end{table*}

In summary, we have shown that, upon accounting for the heterogeneity
of the network and its nonlinear mechanical properties, as well as
microscopic capillary forces responsible for network expulsion, a
complex picture of liquid-liquid phase separation (LLPS) within an
elastic network emerges. Specifically, in addition to the well-known
cavitated macroscopic droplets which fully exclude the network, two
other phases are thermodynamically possible: elastically limited
microdroplets at the network pore scale, and network-including
macroscopic droplets. We introduced two dimensionless parameters
governing the relative stability of these morphologies: the
elasto-capillary number $h$ (\Eq{eq:h}) and the permeo-elastic number $p$
(\Eq{eq:p}), and constructed a phase diagram in the $(p,h)$ plane
(Figs.~\ref{fig:PD}, \ref{fig:NH}G and \ref{fig:stiffening}C) that quantifies the
equilibrium droplet size and network deformation behavior.

Finally, we discuss the relevance of the predicted phases for
experimental systems by providing the order-of-magnitude estimates for
the relevant parameters, presented in Table~\ref{tab:exps}.  For
fluorinated oil demixing in silicone gels (system I with
$h \gg \alpha$), consistently with experimental
observations~\cite{style_liquid-liquid_2018,rosowski_elastic_2020,kim_extreme_2020},
only macroscopic phase separation appears to be relevant: these
networks are too homogeneous, and the surface tension too high, to
permit microphase separation. We note that an independent study
proposes that a combination of mesh size heterogeneity, strongly
heterogeneous nucleation at sparse loci, and network fracture under
stress could lead to the coexistence of microdroplets and cavitated
droplets in these systems~\cite{vidal-henriquez_cavitation_2021}. In
contrast, for cytoplasmic condensates (system II), low surface
tension, large mesh sizes and stiff filaments make permeation the most
likely scenario, while cavitation appears to be ruled out by our
theory: if droplets exclude the cytoskeleton, they are likely to be
size-selected at the network mesh size.  Finally, in the context of
intracellular phase separation in the nucleoplasm (III), all three
scenarios are plausible. In particular, we predict that
mesh-size-selected microdroplets are possible in chromatin for
biologically relevant parameters. Interestingly, the chromatin mesh
size is well below the optical resolution limit: if such microdroplets
exist, they are likely not to have been fully characterized yet.  For
instance, it was recently proposed that phase-separated condensates
are involved in the activation and repression of gene
transcription~\cite{cho_mediator_2018,sabari_coactivator_2018,treen_regulation_2020}. Our
work suggests that such condensates might be elastically limited by
the mechanisms presented herein.

We note that our key theoretical predictions rely on several important
assumptions. First, we have focused on thermodynamic equilibrium
states, neglecting both the kinetic pathways leading to them such as droplet ripening~\cite{rosowski_elastic_2020-1,rosowski_elastic_2020,vidal-henriquez_theory_2020,zhang_mechanical_2020} and merging~\cite{lee_chromatin_2021} and, in
the case of biological systems, their inherently out-of-equilibrium
nature. Second, we have ignored all elastic interactions between the
droplets, which is justifiable when the typical droplet separations
are much greater than their size. Third, we have neglected all
visco-elastic effects in the network: we
thus considered systems over time scales long enough for phase
separation to complete, yet short enough for the network to retain its
mechanical integrity. Exploring the effects of network-mediated
droplet interactions and kinetic processes would provide additional
insights into the behavior of elastically limited droplets, and is
left for future work.

Our study also suggests new ways to engineer size-controlled
microdroplets through elastic limitation. These could be useful for
nanofabrication, as well as to serve as crucibles for chemical
reactions favored by phase exchange: the very high surface-to-volume
ratio would permit fast exchange between the two phases. The
multi-stage chemical reactions can be guided in structured multi-phase
droplets, such as is the case with the ribosome biogenesis in
nucleoli~\cite{feric_coexisting_2016}, where the internal organization
of phases is dictated by their surface
tensions~\cite{mao_designing_2020}. Finally, we note that while we
have focused on the case of droplets that (partially) expel the
network, our theory predicts that capillary forces are reversed when
$p<0$: in this case, the network facilitates phase separation and
condenses around the droplets. This scenario may be involved in the
formation of heterochromatin domains by phase separation of
HP1a~\cite{strom_phase_2017}.  Such network-droplet attraction could
also couple to the nonlinear mechanics of fiber networks to result in
large-scale
stresses~\cite{ronceray_fiber_2016,ronceray_stress-dependent_2019}.

\vspace{1mm}

\textbf{Acknowledgments.}  
SM, AK and MPH are
supported by
NSF through the Princeton University Materials Research Science and
Engineering Center DMR-2011750. P.R. is supported by the NSF through
the Center for the Physics of Biological Function (PHY-1734030). The
project leading to this publication has received funding from the
``Investissements d'Avenir'' French Government program managed by the
French National Research Agency (ANR-16-CONV-0001) and from Excellence
Initiative of Aix-Marseille University - A*MIDEX.

\vspace{-3mm}

\bibliography{Nanodroplets.bib}

\clearpage \newpage

\onecolumngrid

\appendix

\begin{center} {\centering \bf \large{  Liquid demixing in elastic networks: cavitation, permeation,  or size selection? \\ Supplementary Information}}
\end{center}

\section{Mathematical framework: modeling liquid droplets in an elastic network}
\label{sec:framework}

We first discuss the framework we employ to assess the stability of
each of the three scenarios considered in the main text: \emph{(i)}
cavitation, \emph{(ii)} microdroplets, and \emph{(iii)}
permeation. Throughout this article, we consider a single spherical
droplet of phase-separated liquid, in an infinite elastic medium
representing the network. We thus neglect mechanical interactions
between droplets, mediated by the network; this assumption is valid if
the separation between droplets is much larger than their size
(\emph{i.e.}~when the volume fraction of phase-separated droplets is
small). The stability of each scenario is measured by the difference
$\Delta g$ of free energy per droplet volume, compared to an infinite
droplet of phase separated liquid in the absence of an elastic
network. This penalty is captured in three distinct terms: elastic energy stored in
the network, liquid-liquid surface tension, and wetting energy. The
latter two have closed forms as a function of the droplet size and
inner stretch. The
mathematically non-trivial aspect thus lies in the evaluation of the
elastic energy resulting from the network deformation induced by the
droplet.

We characterize the elastic medium by its \emph{stored energy
  function} $W(\lambda_1,\lambda_2,\lambda_3)$ (which we leave
unspecified for now), where the $\lambda_i$'s correspond to the three
principal stretches. This function corresponds to the elastic energy
density in the undeformed material coordinates. We consider a droplet
of size $r_d$ in a spherically symmetric infinite medium. We write the
equilibrium deformation $r=r(R)$, such that a point at distance $R$
from the droplet center in the initial undeformed state is displaced
to radius $r(R)$ in the deformed state. In this geometry, the
principal stretches are the radial stretch
$\lambda_1(R) = \frac{dr}{dR} \equiv s$ and the hoop stretch
$\lambda_2(R)=\lambda_3(R) = r/R \equiv t$.

\begin{figure}[b]
  \centering
  \includegraphics[width=0.7\columnwidth]{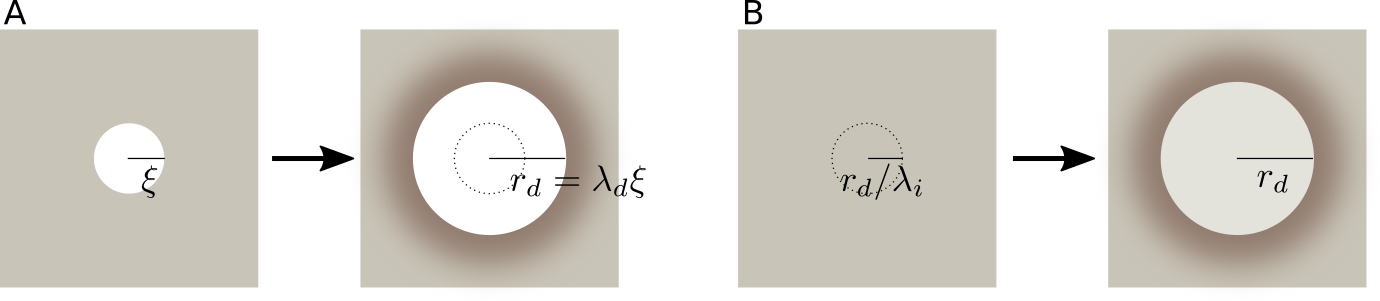}
  \caption{ Geometries of droplets considered here. \textbf{A.}
    Network exclusion, starting from a pore of size $\xi$ stretched by
    a factor $\lambda_d$. \textbf{B.}  Permeation of the droplet
    through the network, with network stretch $\lambda_i$ inside the
    droplet.  }
  \label{fig:geometries}
\end{figure}

We distinguish two geometries, depending on whether the network is
excluded from the droplet (scenarios \emph{(i-ii)}) or included
(scenario \emph{(iii)}):
\begin{itemize}
\item excluded network (\Fig{fig:geometries}A): the medium is modeled
  as an infinite material with an initial spherical pore of radius
  $\xi$ in the reference configuration (left). A droplet of radius
  $r_d = r(\xi)$ stretches this pore by a factor $\lambda_d = r_d/\xi$
  compared to this reference configuration (right). The elastic energy
  stored in the network outside the droplet is thus:
  \begin{equation}
    \label{eq:Eout}
    E_\mathrm{el, out} =  4\pi \int_\xi^\infty  W\left(\frac{dr}{dR}, \frac{r}{R}, \frac{r}{R} \right)  R^2 dR 
  \end{equation}
  Introducing $u=R/\xi$, the radial stretch $s = dr/dR$ and the hoop
  stretch $t= r/R$, we obtain the following form for the elastic
  energy per unit volume of the droplet:
  \begin{equation}
    \label{eq:Eout_nondim}
    \frac{E_\mathrm{el, out}}{v_d}  =  \frac{3}{\lambda_d^3} \int_1^\infty  W\left( s(u), t(u), t(u) \right)  u^2 du \equiv f_\mathrm{out}(\lambda_d)
  \end{equation}
  where $v_d = \frac{4}{3}\pi r_d^3$ is the droplet
  volume. \Eq{eq:Eout_nondim} should be minimized over the deformation
  field $t(u)$, with boundary condition $t(u=1) = \lambda_d$.
\item included (permeated) network (\Fig{fig:geometries}B): the medium is modeled
  as an intact infinite material, and the pores are considered to be
  infinitesimally small. The droplet of radius $r_d$ is placed at the
  center, and imposes a stress discontinuity at its surface. The
  material inside the droplet is isotropically and homogeneously
  deformed with stretch $\lambda_i$. The material outside the droplet
  is deformed in a similar way as previously, and hence the total elastic
  energy reads
  \begin{equation}
    \label{eq:Etot}
    E_\mathrm{el} = E_\mathrm{el,in} +  E_\mathrm{el,out} =  \frac{4}{3} \pi\left(\frac{r_d}{\lambda_i}\right)^3 W(\lambda_i,\lambda_i,\lambda_i) + 4\pi \int_{r_d / \lambda_i}^\infty  W\left(\frac{dr}{dR}, \frac{r}{R}, \frac{r}{R} \right)  R^2 dR 
  \end{equation}
  which, divided by the droplet volume, is:
  \begin{equation}
    \frac{E_\mathrm{el}}{v_d}  = \frac{1}{\lambda_i^3} W(\lambda_i,\lambda_i,\lambda_i) + f_\mathrm{out}(\lambda_i)
    \label{eq:Etot_nondim}
  \end{equation}
  where $f_\mathrm{out}$ was defined in \Eq{eq:Eout_nondim}. In this
  geometry, the fraction of the network excluded from the droplet is
  $\varphi = 1-\lambda_i^{-3}$, so that the wetting energy reads
  $E_\mathrm{wet} = v_d \sigma_p \lambda_i^{-3}$.
\end{itemize}

We finally recapitulate our definition of the free energy for each of
the three phases considered in this article.
\begin{itemize}
\item Cavitation \emph{(i)}: the only contribution to the free energy
  is the elastic penalty, in the infinite-stretch limit of
  \Eq{eq:Eout_nondim}:
  \begin{equation}
    \label{eq:dgi}
    \Delta g_{(i)} = \lim_{\lambda_d\to\infty}  f_\mathrm{out}(\lambda_d).
  \end{equation}
\item Microdroplets \emph{(ii)}: we combine the elastic energy with
  network exclusion (\Eq{eq:Eout_nondim}) with the surface tension. The free energy is found
  by minimizing over the pore stretch (\emph{i.e.}~over the droplet
  radius):
  \begin{equation}
    \label{eq:dgii}
    \Delta g_{(ii)} = \min_{\lambda_d} \left[  \frac{3\gamma}{\lambda_d \xi}  +    f_\mathrm{out}(\lambda_d) \right]
  \end{equation}
  where $\gamma$ is the surface tension. Note that the minimization
  does not always yield a finite value for $\lambda_d$.
\item Permeation \emph{(iii)}: we combine the elastic energy with
  network inclusion (\Eq{eq:Etot_nondim}) with the wetting energy. The free energy is found
  by minimizing over the pore stretch (\emph{i.e.}~over the excluded fraction of the network):
  \begin{equation}
    \label{eq:dgiii}
    \Delta g_{(iii)} = \min_{\lambda_i}  \left[ \frac{ \sigma_p  +  W(\lambda_i,\lambda_i,\lambda_i)}{\lambda_i^{3}}  + f_\mathrm{out}(\lambda_i) \right]
  \end{equation}
  where $\sigma_p$ is the permeation stress. 
\end{itemize}
The mathematically non-trivial part, in all three scenarios, is the
evaluation of the outer elastic energy density
$f_\mathrm{out}(\lambda)$. We combine two approaches, depending on the
class of materials considered, \emph{i.e.}~on the functional form of
$W$. In the case of neo-Hookean materials, we consider slightly
compressible systems, which allows us to solve for the deformation field
analytically, as discussed in \Sec{sec:NH} (corresponding to the
results presented in Fig.~3 of the main text). For strain-stiffening
materials (Fig.~4 of the main text), such an analytical approach is
not possible, and we resort to a numerical estimation of
$f_\mathrm{out}$, as presented in \Sec{sec:NL}. In all cases, the free
energy minimization over the value of $\lambda$ in \Eqs{eq:dgii} and (\ref{eq:dgiii}) is then performed
numerically.

\section{Analytical treatment of slightly compressible neo-Hookean materials}
\label{sec:NH}

Consider Eq.~(\ref{eq:Eout}), written in terms of arbitrary inner and outer radii $R_{min}$ and $R_{max}$: $E_\mathrm{el, out} = \int_{R_{min}}^{R_{max}} 4 \pi R^2 W(\lambda_1,\lambda_2,\lambda_2)dR$.
%
In mechanical equilibrium, $E_\mathrm{el, out}$ is a minimum. Thus, the equilibrium deformation $r=r(R)$ can be obtained from a variational principle as
\begin{equation}
\frac{ \delta E_\mathrm{el, out}}{\delta r(R)} = 8 \pi R \frac{\partial W}{\partial \lambda_2} - 4 \pi \frac{d}{dR} \left(R^2 \frac{\partial W}{\partial \lambda_1} \right) = 0, \label{eq:eelvar}
\end{equation}
or
\begin{equation}
\frac{d}{dR} \left(R^2 W_1 \right) - 2 R W_2 = 0, \label{eq:eelvar2}
\end{equation}
where $W_i \equiv \partial W/\partial \lambda_i$. [Note that here we assume that the system is compressible.  In an incompressible system, the deformation is explicitly determined from $J=\frac{dr}{dR} \left(\frac{r}{R} \right)^2 = 1 \leftrightarrow \frac{dr}{dR} = \left(\frac{r}{R} \right)^{-2}$.]
It is straightforward to show that $d W_1/dR = W_{11} r''(R) + 2 W_{12} (r'(R)/R - r/R^2)$, where $W_{1j} \equiv \partial^2 W /\partial \lambda_1 \partial \lambda_j$. Upon introducing the hoop and radial stretches as $t=r(R)/R$ and $s(t) = dr/dR$, respectively, it can be shown that $r''(R) = ds/dR = ds/dt \, (s-t)/R$ and $d W_1/dR = W_{11} [ds/dt \, (s-t)/R] + 2 W_{12} [s-t]/R$. Thus, Eq.~(\ref{eq:eelvar2}) becomes
\begin{equation}
W_{11} \frac{ds}{dt} = -2 \left(\frac{W_1-W_2}{s-t} + W_{12} \right). \label{eq:eelvar4}
\end{equation} 

Let us next focus on the following simple form for the stored energy function $W$, corresponding to a slightly compressible neo-Hookean network \cite{biwa_cavitation_2006}:
 \begin{equation}
W(\lambda_1,\lambda_2,\lambda_3) = \frac{G}{2} \left[ \lambda_1^2+\lambda_2^2+\lambda_3^2 -3 - 2 (\lambda_1\lambda_2\lambda_3 -1)+ \beta \left( \lambda_1\lambda_2\lambda_3 -1 \right)^2 \right], \label{eq:W1}
\end{equation} 
with $G$ and $\nu = (1-\beta^{-1})/2$ denoting the shear modulus and Poisson's ratio, respectively. It is straightforward to show that, with this choice for $W$, Eq.~(\ref{eq:eelvar4}) becomes  
\begin{equation}
\left(1+ \beta t^4 \right) \frac{ds}{dt} = -2 \left(1+ \beta s t^3 \right). \label{eq:1}
\end{equation}
The exact solution of Eq.~(\ref{eq:1}) is given by \cite{biwa_cavitation_2006} 
\begin{equation}
s(t) = \frac{C_0-\Psi(t)}{\sqrt{1+ \beta t^4}}, \label{eq:2}
\end{equation}
where $C_0$ denotes an integration constant, and
\begin{equation}
\frac{d\Psi(t)}{dt} = \frac{2}{\sqrt{1+ \beta t^4}} \,\,\,\,\, \leftrightarrow \,\,\,\,\, \Psi(t)=\int_{t_0}^t d \tau \frac{2}{\sqrt{1+ \beta \tau^4}}. \label{eq:3}
\end{equation}

Now, consider the case where we have an initial pore of radius $\xi$ embedded within an infinite elastic, neo-Hookean matrix, and the pore walls are subjected to a constant pressure $p_0$.  Far from the cavity, the matrix remains deformation-free, and hence $\lim_{t \rightarrow 1} s(t)=1$.  From the exact solution we immediately obtain
\begin{equation}
  s_I(t) = \frac{\sqrt{1+ \beta}-\int_{1}^t d \tau \frac{2}{\sqrt{1+ \beta \tau^4}}}{\sqrt{1+ \beta t^4}}.
  \label{eq:s1}
\end{equation}
Now, consider subjecting the boundary of the pore to a  stretch $\lambda$ such that $r(\xi) = \lambda \xi$. The corresponding radial stretch is given by 
\begin{equation}
  s_I(\lambda) \equiv \Delta_I = \frac{\sqrt{1+ \beta}-\int_{1}^{\lambda} d \tau \frac{2}{\sqrt{1+ \beta \tau^4}}}{\sqrt{1+ \beta \lambda^4}}.
  \label{eq:s1l}
\end{equation} 
Now, the pressure $p_0$ required to sustain the deformation is given by
\begin{equation}
  \frac{p_0(\lambda, \beta)}{G} = -\frac{W_1}{G \lambda^2} = 1- \frac{\Delta_I}{\lambda^2} - \beta \left(\Delta_I \lambda^2-1 \right).
  \label{eq:p0}
\end{equation}
We obtain the stored elastic energy as the total work of pressure
forces from the undeformed state:
\begin{equation}
 E_{\mathrm{el,out}}(\lambda)  =4\pi \xi^3 \int_1^\lambda p_0(\lambda', \beta) \lambda^{'2} d\lambda'
\end{equation}
Using the formal calculus software SymPy~\cite{meurer_sympy_2017} to
expand the integral in \Eq{eq:s1l} in powers of $\beta^{-1}$
(\emph{i.e.} a weakly compressible expansion), we finally obtain the
following expression for the elastic energy per droplet volume
$f_\mathrm{out}$ as a function of the droplet stretch
$\lambda = r/\xi$:
\begin{equation}
      \label{eq:fout_expansion}
\begin{split}
  \frac{f_\mathrm{out}(\lambda)}{G} = &+ \frac{5}{2} - \frac{3}{\lambda} - \frac{1}{\lambda^{3}} + \frac{3}{2 \lambda^{4}}  \\
& +  \beta^{-1} \left[  - \frac{3}{40} + \frac{6}{5 \lambda^{3}} - \frac{9}{4 \lambda^{4}} + \frac{6}{5 \lambda^{5}} - \frac{3}{40 \lambda^{8}} \right]  \\
& + \beta^{-2} \left[  \frac{1}{48} - \frac{2}{15 \lambda^{3}} + \frac{9}{80 \lambda^{4}} + \frac{9}{80 \lambda^{8}} - \frac{2}{15 \lambda^{9}} + \frac{1}{48 \lambda^{12}} \right]  \\
& + \beta^{-3} \left[  - \frac{15}{1664} + \frac{14}{325 \lambda^{3}} - \frac{1}{32 \lambda^{4}} - \frac{9}{1600 \lambda^{8}} - \frac{1}{32 \lambda^{12}} + \frac{14}{325 \lambda^{13}} - \frac{15}{1664 \lambda^{16}} \right] \\
& + \beta^{-4} \left[  \frac{21}{4352} - \frac{22}{1105 \lambda^{3}} + \frac{45}{3328 \lambda^{4}} + \frac{1}{640 \lambda^{8}} + \frac{1}{640 \lambda^{12}} + \frac{45}{3328 \lambda^{16}} - \frac{22}{1105 \lambda^{17}} + \frac{21}{4352 \lambda^{20}} \right]  \\
& + \beta^{-5} \left[  - \frac{3}{1024} + \frac{22}{1989 \lambda^{3}} - \frac{63}{8704 \lambda^{4}} - \frac{9}{13312 \lambda^{8}} - \frac{1}{2304 \lambda^{12}} - \frac{9}{13312 \lambda^{16}} - \frac{63}{8704 \lambda^{20}} + \frac{22}{1989 \lambda^{21}} - \frac{3}{1024 \lambda^{24}} \right] \\
  & +  O\left(\beta^{-6}\right)                        
\end{split}
\end{equation}
which we use to assess the stability of each phase, as described in
\Sec{sec:framework}. For $\nu>0$ (corresponding to $\beta > 1$ this
expansion converges rapidly, and the results are essentially unaffected
by including additional terms (in practice, for the results presented in Fig.~3
of the main text, we expand up to $\beta^{-8}$).  In particular, we can
read out the $\lambda\to\infty$ limit, corresponding to the cavitated
free energy (\Eq{eq:dgi}):
\begin{equation}
  \label{eq:dgi_NH}
  \alpha \equiv \frac{\Delta g_{(i)}}{G} = \frac{1}{G} \lim_{\lambda\to\infty}  f_\mathrm{out}(\lambda) = \frac{5}{2} - \frac{3}{40 \beta} + \frac{1}{48 \beta^{2}} - \frac{15}{1664 \beta^{3}} + \frac{21}{4352 \beta^{4}} - \frac{3}{1024 \beta^{5}} + \frac{99}{51200 \beta^{6}} +O\left( \beta^{-7}\right)
\end{equation}
Note that we also have $\alpha = p_0^*/G$, where $p_0^*$ is the
cavitation pressure. As expected, in the limit
$\beta \rightarrow \infty$, $p_0^*/G \rightarrow 5/2$, in agreement
with the classic cavitation result for incompressible neo-Hookean
materials.
For reasonable values of $\nu = 1/3$ ($\beta = 3$) or
$\nu=1/4$ ($\beta = 2$), Eq.~(\ref{eq:dgi_NH}) yields
$p_0^*/G \approx 2.48$ and $2.47$, respectively. Finite
compressibility thus reduces the critical cavitation pressure, albeit
to a rather small degree. We also note that for the special case
$\beta=1$ (corresponding to $\nu=0$),
$p_0^*/G = 2 - \sqrt{2} + 4/\sqrt{\pi} \,\, \Gamma^2(5/4)\approx
2.44$, where $\Gamma(x)$ denotes the Euler gamma function, while the
series approximation in Eq.~(\ref{eq:dgi_NH}) yields
$p_0^*/G \approx 2.44$, in excellent agreement with the exact result.
Cavitation pressures for several representative compressibilities are
listed in {\bf{Table 1}}. 
\begin{table}[h!]
\begin{center}
\begin{tabular}{c c c c c c c c} 
\hline\hline
 $\beta$ & 1  & 2 & 3 & 5 & 10 & 50 & $\infty$ \\ 
 $\nu$  &  0 &  1/4 & 1/3 & 0.4 & 0.45  & 0.49 & 1/2\\
 $ \alpha = p_0^*(\beta)/G$  & 2.439 &  2.467 & 2.477 & 2.486 & 2.493  & 2.4985 & 5/2 \\  
 \hline\hline
\end{tabular}
\caption{Critical cavitation pressure of a finite spherical pore in an infinite, slightly compressible Neo-Hookean matrix at varying compressibilities.}\label{table1}
\end{center}
\end{table}

\section{Limit of metastability of microdroplets in the neo-Hookean model}
\label{sec:metastable}

\begin{figure}[hbt]
  \centering
  \includegraphics[width=\columnwidth]{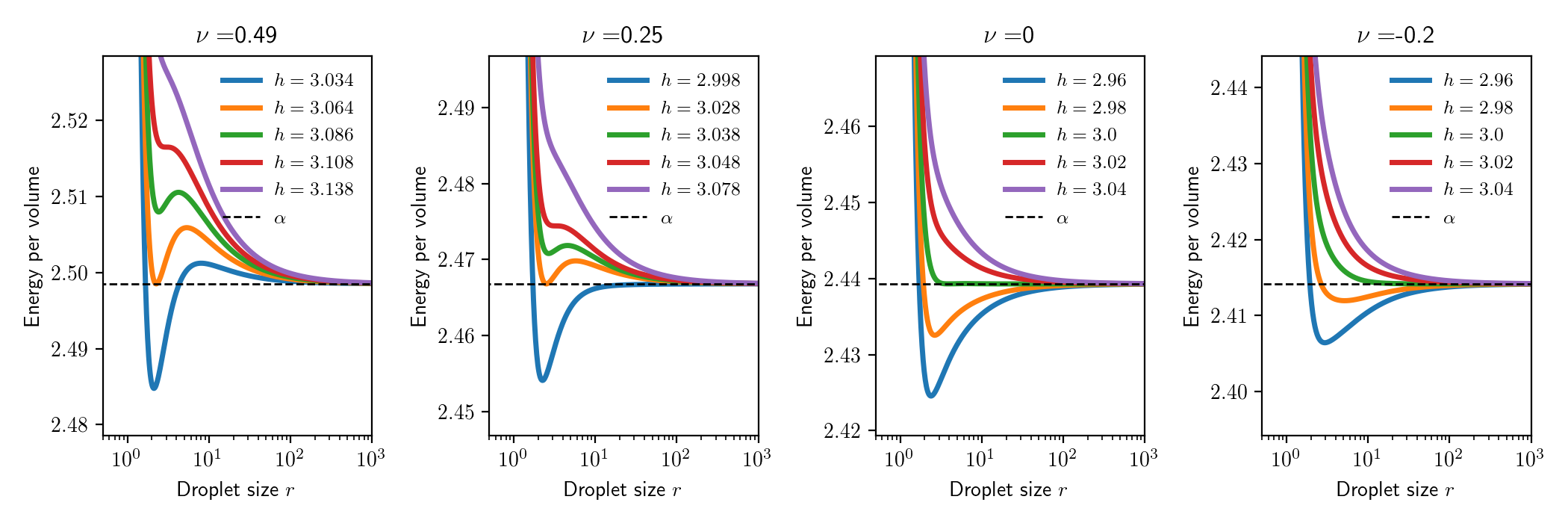}
  \caption{ Plots of $\frac{1}{G} \Delta g (\lambda)$ for different
    values of $h$ and $\nu$. The metastability regimes corresponding
    to a first-order transition are apparent for $\nu >0$ (left two
    panels, with the equilibrium transition $h=h_c$ in orange and the
    end of metastability $h=h^{\dagger}$ in red). For $\nu \leq 0$
    (right two panels), the transition is second-order and occurs at
    $h=3$.  }
  \label{fig:metastable1}
\end{figure}

\begin{figure}[hbt]
  \centering
  \includegraphics[width=0.7\columnwidth]{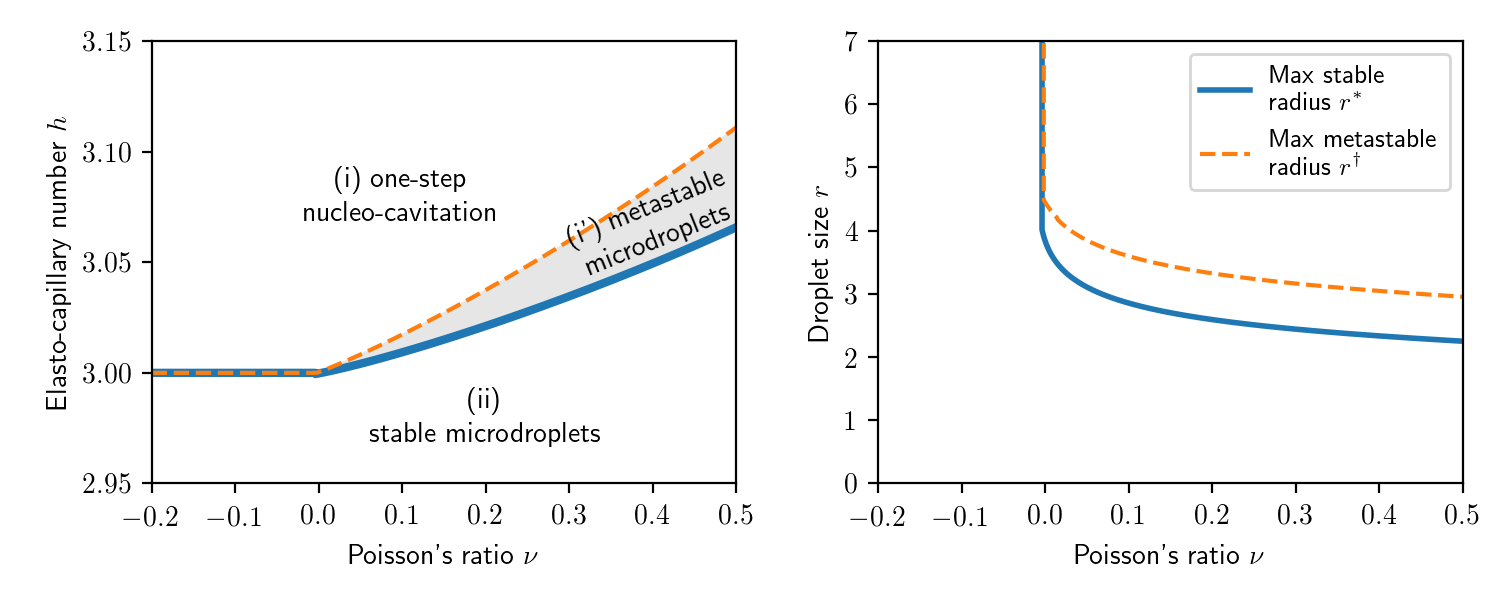}
  \caption{ Left: equilibrium transition line $h_c$ (solid blue line)
    and metastability limit $h^\dagger$ (dashed orange line) as a
    function of Poisson's ratio $\nu$. The shaded area indicates the
    region in which microdroplets can be metastable. Note the very
    limited range of $h$ values represented here. Right: maximum
    radius for stable microdroplets (solid blue line) and metastable
    droplets (dashed orange line). At $\nu<0$ these are infinite, as
    the cavitation transition is continuous.  }
  \label{fig:metastable2}
\end{figure}

We investigate here the nature of the equilibrium transition between
microdroplets (scenario \emph{ii}) and cavitation (scenario \emph{i}),
which is controlled by the elasto-capillary number $h$. To this aim,
it is useful to consider the free energy per volume of a droplet of
size $r = \lambda \xi$ as a function of its stretch $\lambda$, in the
large $\lambda$ limit that can be read out from
\Eq{eq:fout_expansion}:
\begin{equation}
  \label{eq:dgii_metastable}
  \frac{1}{G} \Delta g (\lambda) =  \alpha(\beta) + \frac{h-3}{\lambda} + \frac{A_3(\beta)}{\lambda^3} + O(\lambda^{-4})
\end{equation}
where again $\beta=1/(1-2\nu)$ is the compressibility
parameter, and $A_3(\beta)$ is the coefficient of the inverse cubic term. Interestingly, the leading order in the expansion changes
sign when $h=3$ (independently of $\beta$), and around this value the
free energy is thus dominated by higher-order terms. Depending on the
compressibility, we identify two qualitatively distinct behaviors when
varying $h$, as shown in \Fig{fig:metastable1}:
\begin{itemize}
\item for $\nu > 0$, \emph{i.e.} for usual materials, we have the
  following sequence of regimes:
  \begin{itemize}
  \item $h < 3$: $\Delta g (\lambda)$ exhibits a single minimum at a
    finite $\lambda^*=r^*/\xi$, corresponding to the microdroplets
    scenario. Near equilibrium, droplets larger than $r^*$ would
    shrink (``anti-ripen'') so as to reach the equilibrium size.
  \item $ 3 < h < h_c$: microdroplets are the global free energy
    minimum, but there is a local maximum at $\lambda > \lambda^*$. As
    a result, $\lambda=\infty$ is a local minimum of free energy, and
    cavitated droplets are metastable.
  \item $h_c < h < h^\dagger$: the global minimum of free energy is at
    $\lambda=\infty$, and cavitation is the stable scenario; however,
    a local minimum exists at $r^\dagger$, corresponding to metastable
    microdroplets.
  \item $h>h^\dagger$: the free energy is monotonically decreasing as
    a function of $\lambda$, cavitation is stable and there exists no
    metastable state. 
  \end{itemize}
  The transition between scenarios $(i)$ and $(ii)$ governed by $h$ is
  thus first-order. However, plotting in \Fig{fig:metastable2} the
  values of $h_c$ and $h^\dagger$ over the physical range of Poisson's
  ratio values $\nu$, we note that the range of metastability
  corresponding to this first-order transition is very narrow, and
  restricted to values $3<h<3.11$ for all $\nu$.
\item for $\nu<0$, \emph{i.e.} for auxetic materials, we observe a
  second-order transition between scenarios \emph{(i)} and \emph{(ii)}
  (right panels in \Fig{fig:metastable1}), with a continuous
  divergence of the droplet radius as $r^* \sim (3-h)^{-1/2}$ as
  $h\to 3$.
\end{itemize}
Overall, this analysis shows that at $\nu >0$ the cavitation transition is
\emph{weakly first order}, characterized by the proximity to a
critical point at $\nu=0$, sharp increase of the droplet size near the
transition (as shown in Fig 3C of the main text), and very limited
range of metastability.

\section{Minimal model for strain-stiffening effects}
\label{sec:NL}

The stored energy function $W$ for neo-Hookean
materials~\cite{treloar_physics_2005} in Eq.~(\ref{eq:W1}) does not
capture strain-stiffening effects occuring in macromolecular systems
at large stretches~\cite{ogden_non-linear_1997,
  horgan_molecular-statistical_2002,gent_new_1996,arruda_three-dimensional_1993,boyce_direct_1996}. To
account for such effects, we consider the following modified stored
energy function:
\begin{equation}
W = \frac{G}{2} \left[ (I_1 - 3) - 2(J-1) + \beta (J-1)^2 + 
\left( \frac{I_1-3}{6\, \varepsilon_c}\right)^3
\right], \label{eq:Wss} 
\end{equation} 
where $I_1=\lambda_1^2+\lambda_2^2+\lambda_3^2$ and
$J=\lambda_1 \lambda_2 \lambda_3$, and where $\varepsilon_c$ denotes a characteristic strain at which stiffening effects become
significant. It should be noted that the last term in
Eq.~(\ref{eq:Wss}) emerges as the leading order term in a polynomial
expansion of the classic Arruda-Boyce
\cite{arruda_three-dimensional_1993,boyce_direct_1996} and Gent
\cite{gent_new_1996} models for large-stretch behavior of polymer
systems.

The above choice for $W$ is both convenient and physically-based: (1) At infinitesimally small strains,  
$W\sim G\left[\epsilon_{ij}\epsilon_{ij} + \frac{\nu}{1-2\nu} \epsilon_{ii}\epsilon_{jj}\right]$,
in accordance with linear elasticity theory, where $\beta^{-1}=(1- 2\nu)$ with $\nu$ denoting the Poisson's ratio and $\boldsymbol{\epsilon}$ the linear strain. 
(2) By taking $\varepsilon_c\rightarrow\infty$, we recover the (slightly) compressible neo-Hookean model in Eq.~(\ref{eq:W1}). (3) Asymptotically, $W\sim [(I_1-3)/(6\varepsilon_c)]^3 \sim \left[(\lambda_{chain}^2-1)/(2\varepsilon_c)\right]^3$, indicating a strong stiffening effect when $I_1 \rightarrow (3 + 6\varepsilon_c)$.
Therefore, this form of $W$ can be viewed as a minimal model for slightly compressible, strain-stiffening hyperelastic materials. Specifically, by tuning the parameter $\beta$, we can vary the compressibility with $\beta\rightarrow \infty$ corresponding to a perfectly incompressible material, while by tuning $\varepsilon_c$, we can vary the material  response from weakly stiffening (large $\varepsilon_c$) to strongly stiffening (small $\varepsilon_c$). 

We employ numerical simulations to study the influence of
strain-stiffening of the network on liquid-liquid phase
separation. Specifically, to evaluate $f_\mathrm{out}(\lambda)$ (as
given by \Eq{eq:Eout_nondim}), we discretize the displacement field
over an uneven grid,
$u = [ 1, 1.16, 1.33, 1.51, ..., 28.4, 29.2, u_\mathrm{max} = 30 ]$
(with regular spacing of the values of $\sqrt{u}$). We evaluate the
integral
$\int_1^{u_\mathrm{max}} W\left( s(u), t(u), t(u) \right) u^2 du$
using finite differences of the displacement field, and use the SciPy
optimization package~\cite{virtanen_scipy_2020} (\texttt{scipy.optimize.minimize}) to perform the
multivariate minimization of the energy of the displacement field,
under the constraint $r(u=1) = \lambda$. The outcome of this
optimization is insensitive to the details of the discretization, and
recovers the analytical solution presented in \Sec{sec:NH} in the case
of neo-Hookean materials. We then pipe the resulting function
$f_\mathrm{out}(\lambda)$ into the free energy minimization described
in \Eqs{eq:dgii}-(\ref{eq:dgiii}).

\section{Further discussion of permeation stress $\sigma_p$.}
\label{sec:permeation}

We now discuss the microscopic origin of the permeation stress
$\sigma_p$, and how it could be measured in practice.
This term stems from the difference of wettability between the two
liquids and the network. Microscopically, we can model the filaments
composing the network as cylinders of radius $r_f$, corresponding to a
liquid-solid interface area per unit length of $2\pi r_f$. Taking the
filaments immersed in the majority liquid (liquid 1) as the reference
of energy, the interfacial energy per unit length of a filament
immersed in liquid 2 is thus $2\pi r_f (\gamma_{2S} - \gamma_{1S})$
where $ \gamma_{1S}$ and $\gamma_{2S}$ respectively correspond to the
interfacial energy between the network and liquids 1 and 2 (note that
these interfacial energies are defined at the microscopic level, not
at the network level). Denoting by $\rho$ the volume fraction of the
network in its rest state (we typically consider cases where
$\rho \ll 1$), the liquid-network contact area per unit volume is
$2\rho/r_f$. The difference of energy per unit volume between the
network immersed in liquid 1 and in liquid 2 is thus:
\begin{equation}
  \label{eq:sigmap}
  \sigma_p = \frac{2}{r_f} \rho (\gamma_{2S} - \gamma_{1S}).
\end{equation}
In the presence of strain in the network, its volume fraction may
change: denoting by $\varphi = 1 - 1/(\lambda_1 \lambda_2 \lambda_3)$
the fraction of the network that is expelled compared to the
undeformed state (where the $\lambda_i$'s are principal stretches),
the energy per volume associated with immersing the stretched network
into the minority liquid is thus $(1-\varphi) \sigma_p$, corresponding
to Eq. 5 of the main text. 

At the liquid-liquid interface, the difference in surface energy
results in a capillary force
$F_c \sim 2\pi r_f (\gamma_{2S} - \gamma_{1S})$ on each filament going
through the interface. At the network scale, this implies a stress
discontinuity in the network: the network
is being ``sucked in'' by the best-wetting liquid. Note that while
\Eq{eq:sigmap} relies on microscopic modeling of the network and
applicability of the surface energy at the level of individual
filaments, this stress discontinuity suggests that $\sigma_p$ could
also be measured experimentally, in a way that is independent from
microscopic models. Indeed, consider a tube separating two chambers
containing respectively liquids 1 and 2, with a cork of clamped
network in the tube.  Then $\sigma_p$ corresponds to the pressure
difference one must impose between the two chambers, so that the
liquid-liquid interface remains steady within the network cork. This
provides an experimentally viable way to measure $\sigma_p$. The
existence and microscopic origin of this term was previously noted
by de Gennes in the context of non-deformable porous
media~\cite{de_gennes_liquid-liquid_1984}.

Note that in addition to the difference of liquid-solid surface
energy, it is possible that the rest state of the network changes when
immersed in liquid 2 -- either swelling or shrinking -- and thus that
permeation induces a spontaneous strain in the network. This
qualitatively distinct effect has, in practice, consequences similar
to the effect discussed above, and thus simply results in a
modification of $\sigma_p$ for our purposes.

\section{Estimation of physical parameters.}
\label{sec:exps}

Here we discuss how we obtain the experimental values of parameters
presented in Table I of the main text, for each of the three classes
of systems considered. We focus on obtaining order-of-magnitude
estimates for the two dimensionless parameters introduced in the main
text: the elasto-capillary number $h \equiv 3\gamma /\xi G$ with $\gamma$ the
liquid-liquid surface tension, $\xi$ the network pore size, and $G$
the network shear modulus; and the permeo-elastic number
$p \equiv \sigma_p/G$ with $\sigma_p$ the permeation stress. Note that values
of $\sigma_p$ have not been reported in the literature, to the best of
our knowledge; for this reason, we employ \Eq{eq:sigmap} with typical
values for surface tensions to get order-of-magnitude estimates of its
range of variation. To this end, we substitute
$\rho \approx r_f^2/\xi^2$, with $r_f$ denoting the radius of the filaments
constituting the network. Below, we consider three distinct systems, one synthetic, and two biological ones. 

\textbf{System I} comprises the demixing of fluorinated oil embedded in a
silicone gel, studied in great detail in
Refs.~\cite{style_liquid-liquid_2018,rosowski_elastic_2020,kim_extreme_2020}. The
elastic modulus $G$ is in the range $1.4-280 \si{\kilo\pascal}$ (we
employ a Poisson ratio $\nu = 0.5$ to convert from reported values of
the Young's modulus). Following Ref.~\cite{rosowski_elastic_2020}, we
relate the modulus of this polymer network to its mesh size $\xi$
through $\xi\sim (k_B T / G)^{1/3}$ with
$k_B T=4\times 10^{-21} \si{\joule}$ the thermal energy. Hence,
$\xi \sim 2.4-14 \si{\nano\metre}$. As for the surface tension, we employ $\gamma \approx 4.4 \si{\milli \newton \per \metre}$, as reported in
Ref.~\cite{kim_extreme_2020}. We take a representative molecular radius
$r_f\approx 0.2\si{\nano\metre}$ for PDMS chains, which yields
$\sigma_p \sim 9-300 \si{\kilo \pascal}$ (we emphasize that this is a
rough estimate). The value $\alpha \approx 1.5$ (as the ratio between
cavitation pressure and shear modulus) is reported in Ref.~\cite{style_liquid-liquid_2018}.
We conclude that the range of variation of dimensionless parameters
$h$ for system I is $h \sim 20-700 \gg \alpha$ and
$p \sim 1.1-6.5 \gtrapprox \alpha$, where larger values of $h$ and $p$
both correspond to softer gels. Our theory thus predicts that the
relevant regime is predominantly cavitation (scenario \emph{i}), with
permeation (\emph{iii}) being marginally possible for very stiff
gels. This is consistent with the experimental observation of large,
micron-sized droplets (while the mesh size is in the nanometer range)
that fully exclude the surrounding network, as characterized by
coherent anti-Stokes Raman scattering \cite{kim_extreme_2020}.

\textbf{System III} generally encompasses liquid condensates found in
the nucleus of eukaryotic cells and mechanically interacting with the
chromatin network, both native (such as
nucleoli~\cite{feric_coexisting_2016}) and biomimetic (such as CasDrop
optogenetically activated condensates~\cite{shin_liquid_2018}). Due to
the broad class of systems considered and to the scarcity of available quantitative
data for physical parameters, we report only conservative ranges for
our estimates. Following Ref.~\cite{shin_liquid_2018}, we estimate the
elastic modulus to be in the range $G \sim 10-1000 \si{\pascal}$ and a
mesh size $\xi \sim 7 - 20 \si{\nano \meter}$, with larger mesh sizes
corresponding to softer chromatin. We estimate the surface tension to
be in the range
$\gamma \sim 10^{-7} - 10^{-6} \si{\newton\per\metre}$. Indeed, such
low values of surface tension have been reported for nucleolar
proteins,
$\gamma \sim 4\times
10^{-7}\si{\newton\per\metre}$~\cite{feric_coexisting_2016}. We take a
radius $r_f = 1\si{\nano\metre}$ for DNA, and a volume fraction
$\rho \sim 0.1-0.4$~\cite{shin_liquid_2018}, which yields
$\sigma_p \sim \pm 10 - 100 \si{\pascal}$ (note that the sign of
$\sigma_p$ depends on whether the nucleoplasm or the liquid condensate better wets the chromatin, which is not known \emph{a
  priori}). No value of $\alpha$ has been reported to our knowledge,
and so we take $\alpha \sim 2.5$, corresponding to the neo-Hookean
case, as a default. This results in a very broad range of possible
values for dimensionless parameters, $h\sim 10^{-2} - 10$ and
$p\sim \pm 10^{-2} - 10$. In particular, all three scenarios appear to
be plausible: cavitation (\emph{i}) in soft chromatin and for rather
large values of the surface tension; nanodroplets confined at the mesh
size (\emph{ii}) if chromatin is stiffer and for low liquid-liquid
surface tension; and finally permeation (\emph{iii}) if the
interfacial energy between chromatin and the condensate is low.
Interestingly, only scenario \emph{(i)} has been characterized yet:
both nucleoli and engineered condensates form micron-sized droplets
that have been shown to exclude the surrounding chromatin as they
grow~\cite{shin_liquid_2018}. However, it is possible that
mesh-size-level droplets actually exist, but have not been
characterized yet as they would be significantly below optical
resolution.

\textbf{System II}, finally, encompasses cytoplasmic liquid
condensates such as stress granules and P-bodies, which interact
mechanically with cytoskeletal networks, in particular the actin
cortex.  The main changes compared to system II are the properties of
the elastic network. Reported values for the shear modulus of the
cytoskeleton in intracellular conditions are similar in range to the
nucleus, $G\sim
10-100\si{\pascal}$~\cite{pegoraro_mechanical_2017}. However, the mesh
size of the actin cortex,
$\xi \sim 50-150
\si{\nano\metre}$~\cite{hohmann_cytoskeleton_2019}, is much
larger than that of chromatin, as it is composed of sparser, stiffer
filaments. We take a radius $r_f \sim 2.5 \si{\nano\metre}$ for F-actin
filaments. Ref.~\cite{brangwynne_germline_2009} reports a surface
tension $\gamma \approx 1 \si{\micro \newton \per\metre}$ for
cytoplasmic P-granules. The permeation stress is thus
$\sigma_p \sim 0.2-2 \si{\pascal}$. The range for dimensionless
parameters is thus $h\sim 0.2 - 6$ and $p \sim \pm 10^{-3} - 0.2$.
Interestingly, this excludes cavitation (\emph{i}): permeation
(\emph{iii}) is the predominant scenario, while microdroplets
(\emph{ii}) remain marginally possible. It is therefore an open
question whether permeation actually occurs in experiments.

\end{document}